\documentclass[useAMS,usenatbib,fleqn]{mn2e}

\usepackage[draft]{hyperref}
\usepackage{amssymb}
\usepackage{amsmath}
\usepackage{mathtools}
\usepackage{enumitem}
\usepackage{graphicx}

\usepackage{nicefrac}
\usepackage{bigints}

\title[Estimating dust distances to Type Ia supernovae]{Estimating dust distances to Type Ia supernovae from colour excess time-evolution}
\author[M. Bulla et al.]{M.~Bulla,\thanks{E-mail: mattia.bulla@fysik.su.se} A. Goobar, R.~Amanullah, U.~Feindt and R.~Ferretti \\
Oskar Klein Centre, Department of Physics, Stockholm University, SE 106 91 Stockholm, Sweden
}

\usepackage{color}
\newcommand{\revised}[1]{\textcolor{black}{#1}}
\newcommand{\referee}[1]{\textcolor{black}{#1}}
\newcommand{\refereenew}[1]{\textcolor{black}{#1}}
\newcommand{\refereelast}[1]{\textcolor{black}{#1}}
\newcommand{\ebv}{\mbox{$E(B-V)$}}
\newcommand{\ebvdlos}{\mbox{$E(B-V)_\mathrm{DLOS}$}}
\newcommand{\lmc}{\mbox{LMC$-$}}
\newcommand{\mw}{\mbox{MW$-$}}

\date{Accepted 2017 August 31. Received 2017 August 31; in original form 2017 July 3.}

\begin{document}

\maketitle 

\begin{abstract}
We present a new technique to infer dust locations towards reddened Type Ia supernovae and to help discriminate between an interstellar and a circumstellar origin for the observed extinction. Using Monte Carlo simulations, we show that the time-evolution of the light-curve shape and especially of the colour excess \ebv~places strong constraints on the distance between dust and the supernova. We apply our approach to two highly-reddened Type Ia supernovae for which dust distance estimates are available in the literature: SN~2006X and SN~2014J. For the former, we obtain a time-variable \ebv~and from this derive a distance of \revised{\refereenew{$27.5^{+9.0}_{-4.9}$}} or \revised{\refereenew{$22.1^{+6.0}_{-3.8}$}}~pc depending on whether dust properties typical of the Large Magellanic Cloud (LMC) or the Milky Way (MW) are used. For the latter, instead, we obtain a constant \ebv~consistent with dust at distances larger than \revised{$\sim$~\referee{50} and \refereenew{38}~pc} for \lmc~and \mw type dust, respectively. Values thus extracted are in excellent agreement with previous estimates for the two supernovae. Our findings suggest that dust responsible for the extinction towards these supernovae is likely to be located within interstellar clouds. We also discuss how other properties of reddened Type Ia supernovae -- such as their peculiar extinction and polarization behaviour and the detection of variable, blue-shifted sodium features in some of these events -- might be compatible with dust and gas at interstellar-scale distances.
\end{abstract}

\section{Introduction}
\label{sec:introduction}

The use of Type Ia supernovae (SNe~Ia) as distance indicators has led to the breakthrough discovery of the accelerated expansion of the Universe \citep{riess1998,perlmutter1999}. Almost twenty years later, our ability to perform precision cosmology with SNe~Ia is limited by various systematic uncertainties. Among these, one crucial uncertainty is given by our poor knowledge of what causes the colour-brightness relation found in SNe~Ia. Redder SNe~Ia are observed to be typically fainter than bluer ones, but whether this stems from intrinsic SN colour variations, dust reddening or a combination of the two is unclear \citep[see e.g.][]{wang2009,foley2011}.

The effect of dust on the observed reddening is typically described by the total-to-selective extinction ratio \mbox{$R_\mathrm{V}=A_\mathrm{V}/(A_\mathrm{B}-A_\mathrm{V})\equiv A_\mathrm{V}/\ebv$}, where $A_\mathrm{X}$ denotes the extinction in a given band $X$. While $R_\mathrm{V}\sim3$ is found for extinguished stars in our Galaxy \citep{fitzpatrick2007}, significantly lower values are obtained for SNe~Ia both from individual reddened events \citep[$\sim$~1.3$-$1.8,][]{krisciunas2006,eliasrosa2006,eliasrosa2008,wang2008a,folatelli2010,amanullah2014,amanullah2015} and larger samples \citep[$\lesssim$~2,][]{tripp1998,astier2006,nobili2008,kowalski2008,burns2014}. Similar low-$R_\mathrm{V}$ values are inferred from spectropolarimetric measurements and specifically from the observed correlation between extinction and wavelength of maximum polarization $\lambda_\mathrm{max}$, $R_\mathrm{V}\sim5.5~\lambda_\mathrm{max}$ \citep{serkowski1975,whittet1978}. While $\lambda_\mathrm{max}\sim0.55~\mu$m and thus $R_\mathrm{V}\sim3$ are typical values for Galactic stars, polarization spectra of some highly-reddened SNe~Ia are characterized by $0.2\lesssim\lambda_\mathrm{max}\lesssim0.4~\mu$m and thus $1.1\lesssim R_\mathrm{V}\lesssim$~2.2 \citep{patat2015,zelaya2017}.

One possible interpretation for the low-$R_\mathrm{V}$ towards some SNe~Ia is that they arise from interstellar (IS) dust characterized by smaller grains than those in our Galaxy \citep[see e.g.][]{gao2015,nozawa2016,hoang2017}. In particular, \citet{phillips2013} analysed spectra of 32~reddened SNe~Ia and found a good correlation between extinction and strength of a diffuse interstellar band at 5780~\AA, which they interpret as an evidence that the bulk of the observed reddening stems from IS clouds. The reason why the IS environment around SNe~Ia appears to prefer small grains is still puzzling. However, \citet{hoang2017} recently suggested that a shift to smaller sizes might be the consequence of cloud-cloud collisions induced by the SN radiation pressure. 
 
An alternative explanation for low-$R_\mathrm{V}$ values is that they arise from circumstellar (CS) dust \citep{wang2005,patat2006,goobar2008}. In particular, Monte Carlo calculations by \citet{goobar2008} showed that the wavelength-dependence of scattering in CS dust can yield $R_\mathrm{V}$ values in the range between $\sim$~1.5 and 2.5, in good agreement with observations. This interpretation would also be consistent with the presence of CS gas inferred for some SNe~Ia from the observation of variable, blue-shifted sodium and potassium spectral features \citep[][but see also \citealt{chugai2008} for a different interpretation within the IS scenario]{patat2007,blondin2009,simon2009,sternberg2011,
foley2012,maguire2013,graham2015,ferretti2016}.

\referee{An interesting approach to estimate the dust location towards SNe is through the detection of light scattered by dust. These so-called light echoes were detected for a handful of SNe~Ia and interpreted as arising in IS clouds \citep{schmidt1994,sparks1999,cappellaro2001,
quinn2006,wang2008b,crotts2008,drozdov2015,crotts2015,yang2017}. 
Light echo measurements, however, are usually challenging as they require high-resolution (typically space-based), late-time imaging observations to distinguish contributions of scattered from direct light.}




Here we present a new technique to infer dust locations in reddened SNe~Ia and help discriminate between an IS and a CS origin of extinction. This is based on the peculiar effect of dust on the observed light-curve shape \citep{amanullah2011} and particularly on the \ebv~evolution with time. Details of our approach and of its implementation in Monte Carlo simulations are first described in Section~\ref{sec:simulations}. In Section~\ref{sec:observations}, we test our technique and estimate dust locations for two well-studied highly-reddened SNe~Ia for which dust distance estimates are available in the literature. Finally, we compare our findings to previous studies in Section~\ref{sec:discussion} and draw conclusions in Section~\ref{sec:conclusions}.

\section{Simulations}
\label{sec:simulations}

In this section, we present results of our simulations. After a brief description of our Monte Carlo code (Section~\ref{sec:models}), we
study how \revised{photons} scattered by dust into the direct line-of-sight (DLOS) affect SN light-curve properties and provide constraints on dust location (Section~\ref{sec:lc}). In particular, we will first address the impact on the light-curve width (Section~\ref{sec:lcbroad}) and then on the colour excess as a function of time (Section~\ref{sec:predictions}). 

\begin{figure*}
\begin{center}
\includegraphics[width=0.85\textwidth]{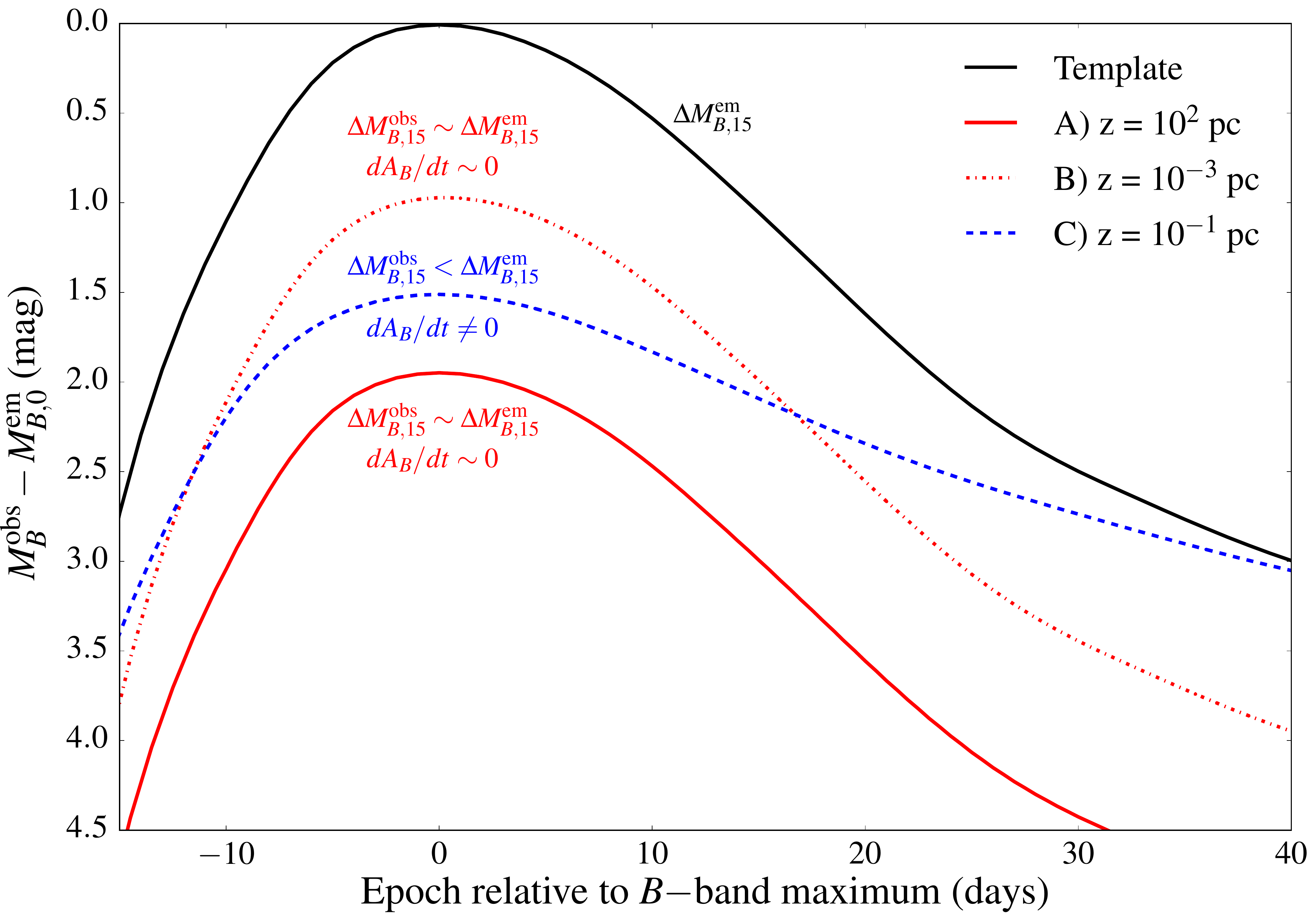}
\caption{Effect of scattered \revised{photons} on the $B-$band light-curve shape. A template $B-$band light curve, characterized by $\Delta M^\mathrm{em}_{B,15}$, is shown in black. Light curves reddened with \lmc type dust and \ebvdlos$=0.5$~mag are shown for three different distances: $z=10^{2}$~pc (solid red line), $z=10^{-3}$~pc (dot-dashed red line), $z=10^{-1}$~pc (dashed blue line). }
\label{fig:lc}
\end{center}
\end{figure*}

\subsection{Models}
\label{sec:models}

We simulate the propagation of SN radiation in a dusty medium using a Monte Carlo code similar to those described by \citet{goobar2008} \revised{and \citet{amanullah2011}}\footnote{\revised{The code used in \citet{amanullah2011} is available at \url{https://github.com/rahmanamanullah/csdust} .}}. A number $N_\textrm{ph}$ of Monte Carlo \revised{photons} are created at the origin and launched into the computational domain assuming isotropic emission. Each \revised{photon} travels unperturbed until it reaches the dust region inner boundary (see below), at which point a random number $\xi_1\in[0,1)$ is drawn and the optical depth calculated as $\tau=-\log(1-\xi_1)$. The physical path length travelled by the \revised{photon, $ds$,} is then calculated from
\begin{equation}
\tau = \int n\,(\sigma_\mathrm{s}+\sigma_\mathrm{a})\,ds~,
\end{equation}
where $n$ is the number density of dust particles, and $\sigma_\mathrm{s}$ and $\sigma_\mathrm{a}$ are the scattering and absorption cross sections, respectively. Upon reaching the interaction point, a \revised{photon} is either absorbed or scattered depending on the dust albedo
\begin{equation}
a = \frac{\sigma_\mathrm{s}}{\sigma_\mathrm{s}+\sigma_\mathrm{a}}~.
\end{equation}
Specifically, a second random number $\xi_2\in[0,1)$ is drawn and an absorption event selected when $\xi_2>a$, while a scattering event when $\xi_2\le a$. In the former case, \referee{the absorbed photon is removed from the simulation and it does not contribute to the extracted light curve (see below)}; in the latter case, a new direction is chosen by sampling a scattering angle $\Theta$ from the Henyey-Greenstein function,
\begin{equation}
\cos\Theta = \frac{1}{2g}\Bigg[1+g^2-\bigg(\frac{1-g^2}{1-g+2\,g\,\xi_3}\bigg)^2\Bigg] ~,
\end{equation} 
where $\xi_3$ is a third random number chosen in the [0,1) interval and $g=\langle\cos\Theta\rangle$ \citep{henyey1941,witt1977}. The \revised{photon} is then moved to the next interaction point and this procedure repeated until the \revised{photon} is either absorbed or leaves the dust region. In the latter case, the escaping \revised{photon} is saved and assigned (i) a final direction and (ii) a time delay with respect to a non-interacting \revised{photon} emitted along the DLOS, $\Delta t$, calculated as in \citet{amanullah2011}.

Following \citet{amanullah2011}, the dust region is modelled as a spherical shell with outer radius $r_\mathrm{out}$ and inner radius $r_\mathrm{inn}=0.95~r_\mathrm{out}$. In this work, $z=r_\mathrm{inn}$ will be taken as representative for the distance between the SN and the dust region. For small distances ($z\lesssim$~0.1~pc), this shell can be thought as representative of a spherical CS shell; for larger distances ($z\gtrsim$~1~pc), it provides a good approximation to an IS dust slab\footnote{For a \revised{photon} scattered by dust into the DLOS, the path travelled is longer -- and thus the time delay larger -- in a slab compared to a spherical-shell case. However, for $z\gtrsim$~1~pc and for scattered \revised{photons} reaching the observer at epochs considered here ($t\lesssim70$~d since explosion), this time difference can be neglected as it is typically much smaller 
than $\Delta t$.} at distance $z$. Wavelength-dependent dust properties ($a$, $\sigma_\mathrm{a}$ and $g$) used in this work correspond to \lmc~and \mw type dust. These are taken from \citet{weingartner2001} and \citet{draine2003}, respectively, and are the same as in table~1 of \citet{goobar2008}. The number density $n$ is assumed to be constant throughout the dust region, and its value set to give a desired colour excess \ebvdlos~ along the DLOS \citep{goobar2008}. \referee{In the case of a CS spherical shell, the assumed constant density implies that the total dust mass $M_\textrm{dust}$ increases with dust distance. Specifically, we find
\begin{equation}
\frac{M_\textrm{dust}}{M_\odot} = W~\bigg[\frac{\ebvdlos}{1~\textrm{mag}}\bigg]~\bigg[\frac{r_\mathrm{inn}}{0.01~\textrm{pc}}\bigg]^2~~,
\end{equation}
where $W=9.5\times 10^{-4}$ and $8.0\times 10^{-4}$ for \lmc~and \mw type dust, respectively. \refereelast{This equation is derived assuming a spherical shell for the dust region and thus the estimated mass is representative of a CS shell only (i.e. dust at small distances). In contrast, it does not apply to a cloud at large distances that should instead be thought as an IS dust slab with a given depth (see above).}}

\revised{Our Monte Carlo code is publicly available and can be found at \url{https://github.com/mbulla/dust3d} .}

\subsection{Impact of scattered \revised{photons} on the light curve}
\label{sec:lc}

\revised{Photons} leaving the dust region and escaping to the observer are collected and used to construct multi-band light curves. \revised{As in \citet{amanullah2011}, pristine SN~Ia light curves are constructed from the \citet{hsiao2007} spectral templates}, which provides us with UBVRI fluxes from $-$20 to $+$85~d relative to $B-$band maximum. Using these, the SN emitted magnitude in a given band $X$ can be expressed as
\begin{equation}
M^\textrm{em}_{X,\textrm{t}} = - 2.5\,\log\,f^\textrm{em}_X(t)\,+\revised{C_X}~,
\end{equation}
where $f^\textrm{em}_X(t)$ is the flux emitted in the $X-$band at time $t$ since explosion, and \revised{$C_X$} some constant. At each given time $t$, the observed magnitude in the $X-$band is then calculated summing over all the escaping \revised{photons} and taking into account their time delays $\Delta t$ (see Section~\ref{sec:models}):
\begin{equation}
M^\textrm{obs}_{X,\textrm{t}} = - 2.5\,\log\frac{\sum_i  f^\textrm{em}_X(t-\Delta t^i)}{N_\textrm{ph}}\,+\revised{C_X}~.
\end{equation}
Finally, the extinction $A_X(t)$ can be easily derived as
\begin{equation}
A_{X,\textrm{t}} = M^\textrm{obs}_{X,\textrm{t}} - M^\textrm{em}_{X,\textrm{t}} = - 2.5\,\log\frac{\sum_i  f^\textrm{em}_X(t-\Delta t^i)}{N_\textrm{ph}\,f^\textrm{em}_X(t)}~.
\end{equation}
In the absence of dust extinction, for instance, \textit{all} the emitted \revised{photons} reaching the observer would have $\Delta t = 0$, hence \mbox{$M^\textrm{obs}_{X,\textrm{t}} = M^\textrm{em}_{X,\textrm{t}}$} and $A_{X,\textrm{t}} = 0$. 

Following the work done by \citet{amanullah2011}, here we investigate the impact of dust on the observed light curve. In particular, we choose to describe the light-curve shape by means of two parameters: (i) the widely-used decline rate in the $B-$band, \mbox{$\Delta M^\mathrm{obs}_{B,15}=M^\textrm{obs}_{B,15} - M^\textrm{obs}_{B,0}$}, and (ii)  the observed colour excess \ebv~$=A_B - A_V$ relative to the pristine light curve. Fig.~\ref{fig:lc} shows $B-$band light curves for dust characterized by \ebvdlos~$=0.5$~mag and located at three different distances: 0.001, 0.1 and 100~pc. As already pointed out by \citet{amanullah2011}, different dust locations translate into different light-curve shapes:
\begin{itemize}[leftmargin=*]
\item At very large distances ($z=100$~pc), the relative contribution of \revised{photons} scattered into the DLOS and reaching the observer at epochs considered in this work ($t\lesssim70$~d since explosion) is negligible. Compared to the pristine light curve, the observed light curve is therefore fainter because of dust extinction but characterized by a similar shape, i.e. $\Delta M^\mathrm{obs}_{B,15}\sim\Delta M^\mathrm{em}_{B,15}$ and $A_B$, $A_V$ and thus \ebv~are constant with time.
\item At very small distances ($z=0.001$~pc), the relative contribution of \revised{photons} scattered into the DLOS is significant and hence the observed light curve is brighter compared to the very-distant dust case described above. However, scattered \revised{photons} arrive almost simultaneously with non-interacting \revised{photons} ($\Delta t \sim 0$) and thus the light-curve shape is again similar to the pristine one.
\item At intermediate distances ($z=0.1$~pc), the contribution of \revised{photons} scattered into the DLOS is relatively large and their time delay of the order of days. Therefore, \revised{photons} reaching the observer at a given time $t$ were emitted at different times and with different luminosities. This affects the observed light-curve shape significantly and translates into an observed broadening after maximum light ($\Delta M^\mathrm{obs}_{B,15}<\Delta M^\mathrm{em}_{B,15}$). \referee{An effect of this light-curve broadening is to make $A_B$, $A_V$ and thus the colour excess \ebv~appear as time-variable}.
\end{itemize}
\referee{As clearly seen at intermediate distances, we note that the extra-flux from scattered packets shapes the light curve not only decreasing $\Delta M_{B,15}$ but also increasing the rise time to peak brightness. This effect was already shown and extensively discussed by \citet[][see section 6.3]{amanullah2011}.}

In the following, we investigate the two effects just discussed more closely and study how $\Delta M^\mathrm{obs}_{B,15}$ (Section~\ref{sec:lcbroad}) and \ebv~(Section~\ref{sec:predictions}) are affected by dust at different locations and with different \ebvdlos~values. In particular, we focus on how these two parameters can be used to place constraints on the dust location towards SNe~Ia. 

\subsubsection{Light-curve broadening}
\label{sec:lcbroad}

\begin{figure}
\begin{center}
\includegraphics[width=1\columnwidth]{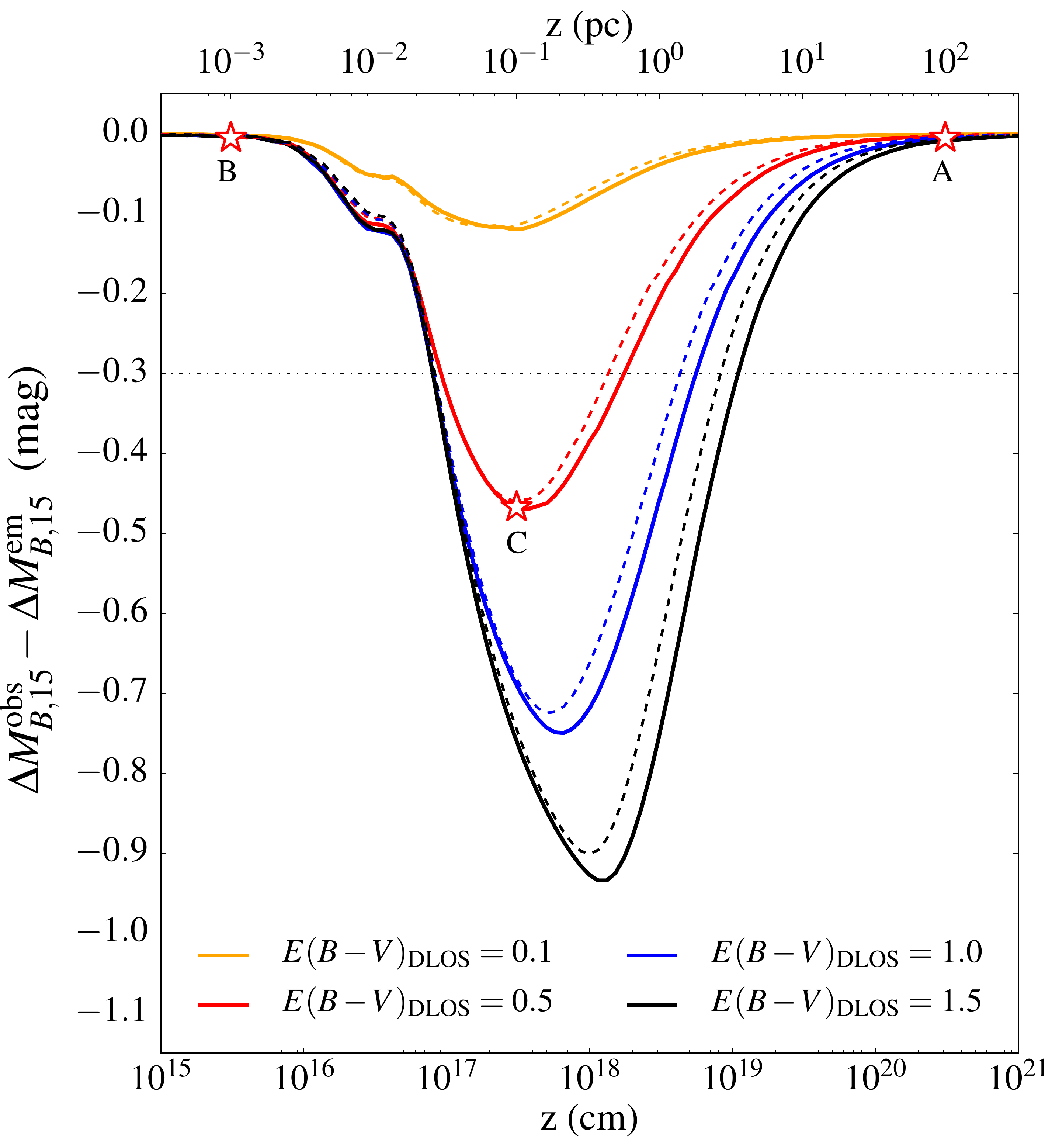}
\caption{Light-curve broadening ($\Delta M^\mathrm{obs}_{B,15}-\Delta M^\mathrm{em}_{B,15}$) as a function of dust distance $z$ for different \ebvdlos~values. Solid lines refer to \lmc type dust, while dashed lines to \mw type dust. The horizontal dot-dashed line marks the light-curve broadening for a SN Ia with $\Delta M^\mathrm{em}_{B,15}=1.1$~mag broadened to and observed as a slow-evolving SN Ia with $\Delta M^\mathrm{obs}_{B,15}=0.8$~mag (see text for explanation). For \ebvdlos~$=0.5$~mag (red lines), the three white stars highlight values corresponding to the A, B and C light curves of Fig.~\ref{fig:lc}.}
\label{fig:dm15}
\end{center}
\end{figure}

Fig.~\ref{fig:dm15} shows the effect of light-curve broadening as a function of dust distance $z$ for different compositions and extinction values. Overall, the general behaviour in light-curve broadening is qualitatively similar between the different models: little broadening is observed for nearby ($z\lesssim0.001$~pc) and fairly distant ($z\gtrsim100$~pc) dust; larger broadening is observed for intermediate distances, with the exact extent depending on both dust composition and extinction.

First, we notice that the difference in dust composition is only marginally affecting the light-curve broadening, with \lmc~and \mw type dust models predicting similar $\Delta M^\mathrm{obs}_{B,15}$ values at any dust distance. On the contrary, the dependence of light-curve broadening on the extinction is remarkable. As \ebvdlos~increases, the light-curve broadening becomes more significant and the dust distance at which the effect is maximum (corresponding to the curve minimum in Fig.~\ref{fig:dm15}) shifts towards larger values. While the former effect simply reflects the increasing contribution of scattered \revised{photons}, the latter is a consequence of the increasingly larger distribution of time delays $\Delta t$ of \revised{photons} reaching the observer.

The light-curve broadening for the \ebvdlos=$0.1$ mag dust model is relatively moderate, with the change in $\Delta M_{B,15}$ reaching a maximum of roughly \refereenew{0.1}~mag at \mbox{$\sim\refereenew{0.1}$~pc}. In contrast, larger values of \ebvdlos~leads to a broadening which is significant when dust is located at distances between 0.01 and a few pc. In this distance range, for instance, models with \mbox{\ebvdlos$=0.5$~mag} are affected by changes in $\Delta M_{B,15}$ that vary \revised{from roughly 0.15 to 0.5~mag}. These numbers become even more extreme when going to larger extinction values, with \revised{0.15 to \refereenew{0.7 (0.9})~mag} changes in $\Delta M_{B,15}$ for \ebvdlos$=1.0$ (1.5)~mag. 

\begin{figure*}
\begin{center}
\includegraphics[width=0.98\textwidth]{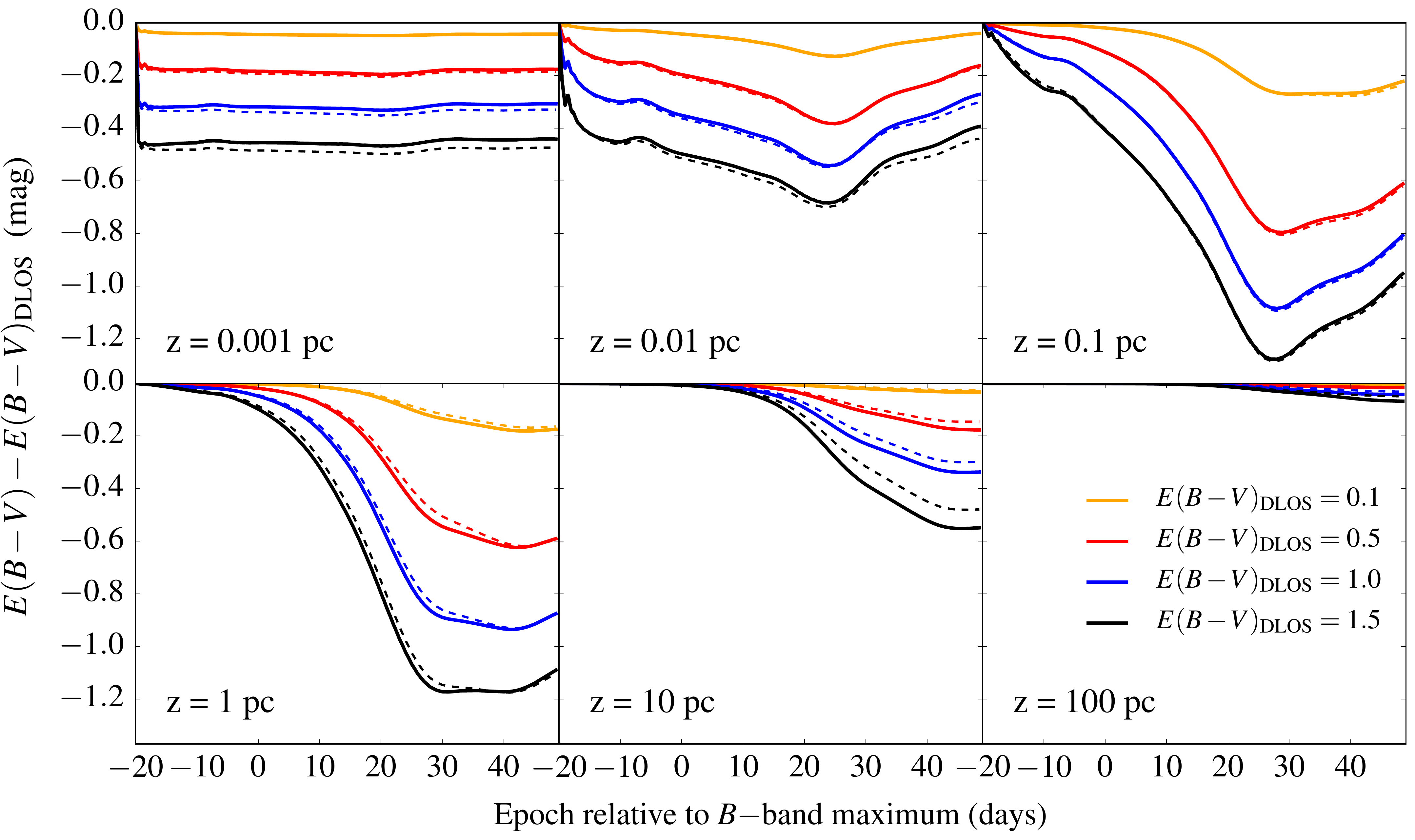}
\caption{Temporal variation of the colour excess \ebv~for different \ebvdlos~values and dust distances. Solid lines refer to \lmc type dust, while dashed lines to \mw type dust. }
\label{fig:ebv}
\end{center}
\end{figure*}

$\Delta M^\mathrm{em}_{B,15}$ values for normal SNe~Ia span a range between roughly 0.8 and 1.5~mag \citep[see e.g.][]{hillebrandt2013}. For highly-extinguished SNe~Ia, a range of intermediate distances ($\sim0.01-1$~pc) can thus be immediately ruled out as they would require extremely large value of $\Delta M^\mathrm{em}_{B,15}$ (i.e. an extremely fast-declining pristine light curve). For instance, the well-studied SN~2006X (see also Section~\ref{sec:2006X}) was estimated to have an observed decline of \mbox{$\Delta M^\mathrm{obs}_{B,15}\sim1.1$~mag} and a colour excess of \mbox{\ebv~$\sim1.5$~mag} \citep{wang2008a,folatelli2010}. In the case reddening along SN~2006X was due
to dust \revised{located between 0.1 and 1~pc from the SN, the pristine light curve would need to have unreasonably large decline rates for normal SNe Ia \mbox{($\Delta M^\mathrm{em}_{B,15}\gtrsim 2$~mag)}}.    
This argument would point -- at least for highly-extinguished SNe~Ia -- towards dust located at either rather small ($z\lesssim0.01$~pc) or large ($z\gtrsim1$~pc) distances. We note, however, that nearby dust at $z\lesssim0.01$~pc is in strong tension with limits inferred from the lack of infrared thermal emission in some normal SNe~Ia \citep[$\gtrsim$~10$^{18}$~cm,][]{johansson2013,johansson2017}. In addition, dust is unlikely to be present at $z\lesssim10^{16}$~cm ($\sim$~0.003~pc) as at these distances it would be destroyed by the SN radiation \citep[see e.g. discussion in][]{amanullah2011}.

Because of the broadening effect described above, the \revised{Hsiao} template will be stretched by a convenient amount when comparing models to data in Section~\ref{sec:observations}. In particular, for a given dust distance, we will choose the specific $\Delta M^\mathrm{em}_{B,15}$ value required to match the SN observed decline rate $\Delta M^\mathrm{obs}_{B,15}\le\Delta M^\mathrm{em}_{B,15}$ after interaction with dust.

\subsubsection{Time-variable colour excess}
\label{sec:predictions}

Fig.~\ref{fig:ebv} shows the temporal variation of the colour excess \ebv~for different dust locations, compositions and extinction values. Overall, the general behaviour of \ebv~with time is similar for all the models and for the scope of our study can be characterized by two main phases: 
\begin{itemize}[leftmargin=*]
\item In a first phase, the relative contribution of \revised{photons} scattered into the DLOS is negligible and the observed flux dominated by non-interacting \revised{photons} with $\Delta t=0$. Each $X-$band light curve is simply dimmed of a constant $A_X$ value, and thus \ebv~is \textit{constant} with time.
\item In a second phase, the relative contribution of \revised{photons} scattered into the DLOS and reaching the observer with some time delay $\Delta t>0$ is significant. Given that scattered \revised{photons} reach the observer at different times, the extinction $A_X$ in each $X-$band is no longer constant with time. In addition, the wavelength-dependence of the scattering and absorption properties causes different time-evolution in different bands, hence \ebv~is also \textit{time-dependent}.
\end{itemize}
Similarly to what is found for $\Delta M^\mathrm{obs}_{B,15}$ (see Section~\ref{sec:lcbroad}), the evolution of \ebv~is weakly dependent on the dust composition, with \lmc~and \mw type dust predicting similar curves. As expected, we notice that the overall change in \ebv~during the second phase is larger for higher values of \ebvdlos. In addition, the transition between the two phases occurs at different epochs depending on the dust location. For very small distances (\mbox{$z\sim0.001$~pc}), all scattered \revised{photons} arrive at the observer with time delays $\Delta t\lesssim1$~d and thus \ebv~starts departing from \ebvdlos~on these short time-scales. As the dust distance increases, the time delays for scattered \revised{photons} becomes larger and thus \ebv~starts deviating from \ebvdlos~on increasingly longer time-scales. In particular, \ebv~starts to deviate from \ebvdlos~already before $B-$band maximum for dust located at $z\lesssim1$~pc, while only post-peak for dust located farther away.

An important aspect shown by Fig.~\ref{fig:ebv} is that  the time evolution of \ebv~is significantly different for dust with different extinctions \ebvdlos~and different locations $z$. Estimating how the \ebv~for reddened SNe~Ia changes with time can thus place strong constraints on the dust location. In the next Section, we will use this model prediction to infer dust locations towards reddened SNe~Ia.



\section{Comparison to observations}
\label{sec:observations}

In this section, we compare the \ebv~evolution predicted by our models with that we calculate for historical highly-reddened SNe~Ia, with the goal of estimating dust distances in these events. In particular, we test our technique by focusing on two well-studied SNe~Ia for which dust distances have been estimated from light echo detections: SN~2006X (Section~\ref{sec:2006X}) and SN~2014J (Section~\ref{sec:2014J}). 

Following \citet{amanullah2014}, \ebv~values for the two SNe are calculated by comparing their observed colours to those of SN~2011fe, a normal ``plain vanilla'' SN~Ia \citep{wheeler2012} we take as representative of a pristine, unreddened SN~Ia\footnote{\revised{We note that colours of SN~2011fe are remarkably similar to those of the Hsiao templates used in our simulations (see Section~\ref{sec:lc}), and this is particularly true for the colours ($B-V$) and epochs ($\gtrsim$~$-10$~d relative to peak) investigated in this work (see figs.~11 and 12 of \citealt{pereira2013}).}} due to its very-low extinction \citep{patat2013}. The errors on the estimated \ebv~values are mostly dominated by the assumed conservative uncertainties on the intrinsic colour scatter in SNe~Ia (0.15~mag in the $B$ and $V$ bands, see \citealt{amanullah2014}). In our study, we choose to average photometric points of each SN in time-windows of 10 (SN~2006X) and 5 (SN~2014J) days in order to predict \ebv~values with uncertainties smaller than 0.05~mag.

\subsection{SN~2006X}
\label{sec:2006X}

SN~2006X was discovered by \citet{suzuki2006} in the M100 galaxy and found to suffer heavy host extinction. In particular, using photometric data, \citet{wang2008a} derived an \textit{averaged} colour excess for SN~2006X of \mbox{\ebv~$=1.42\pm0.04$~mag}. Here we use multi-band photometric data from tables~3 and 4 of \citet{wang2008a} to investigate the evolution of \ebv~with time. Five phase intervals are chosen, and the \ebv~values extracted in each interval are reported in Table~\ref{tab:sn2006x}. We notice that \ebv~is roughly constant and close to 1.5~mag from about $-4$ to $+23$~d relative to maximum, while we observe a drop to \ebv~$\sim$~1.3~mag around $+35$~d.

\begin{figure*}
\begin{center}
\includegraphics[width=1\textwidth]{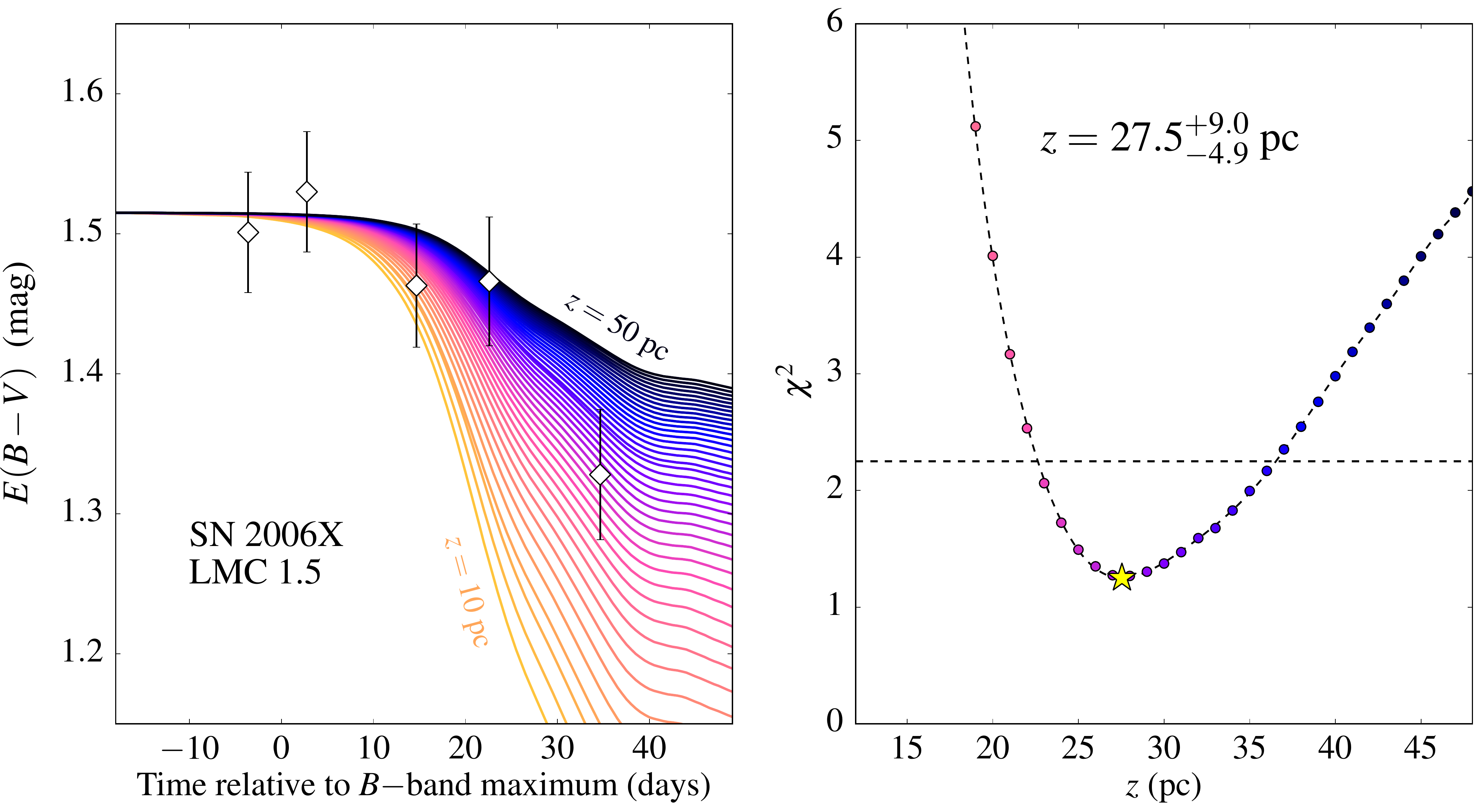}
\caption{\textit{Left panel.} \ebv~temporal evolution of SN~2006X (white diamonds) compared to model predictions for \lmc type dust with \ebvdlos~$=1.5$~mag. Modelled curves are calculated between 10 and 50 pc \referee{(from orange to black)} with a step size of 1 pc, and have been shifted by \referee{+$0.015$~mag to match the averaged value of the first two epochs of SN 2006X}. \textit{Right panel.} $\chi^2$ as a function of dust distances, where each point corresponds to a  curve in the left panel. A \refereenew{12}th-order polynomial fitting (dashed curve) is performed and used to identify the most-probable distance (yellow star) and its uncertainties. The corresponding values are also reported in the panel.}
\label{fig:sn2006X_LMC}
\end{center}
\end{figure*}

Colour excess values thus estimated are shown in the left panel of Fig.~\ref{fig:sn2006X_LMC}, together with curves predicted by a series of \lmc type dust models. These models refer to dust distances from 10 to 50~pc, with a step size of 1~pc. \referee{To overcome our lack of knowledge about the true \ebvdlos~value, we use models with \ebvdlos~$=1.5$~mag but apply a shift of +0.015~mag so that modelled curves at early times are equal to the averaged value of the first two epochs of SN~2006X.}
As we can see, all the selected models predict a post-maximum drop in the \ebv~value compatible with that observed in SN~2006X. To select the most-probable distance, we perform a simple $\chi^2$ test for each modelled curve. As illustrated in the right panel of Fig.~\ref{fig:sn2006X_LMC}, we find a most-probable distance of \revised{\refereenew{$z=27.5^{+9.0}_{-4.9}$}~pc}. 

\begin{figure*}
\begin{center}
\includegraphics[width=1\textwidth]{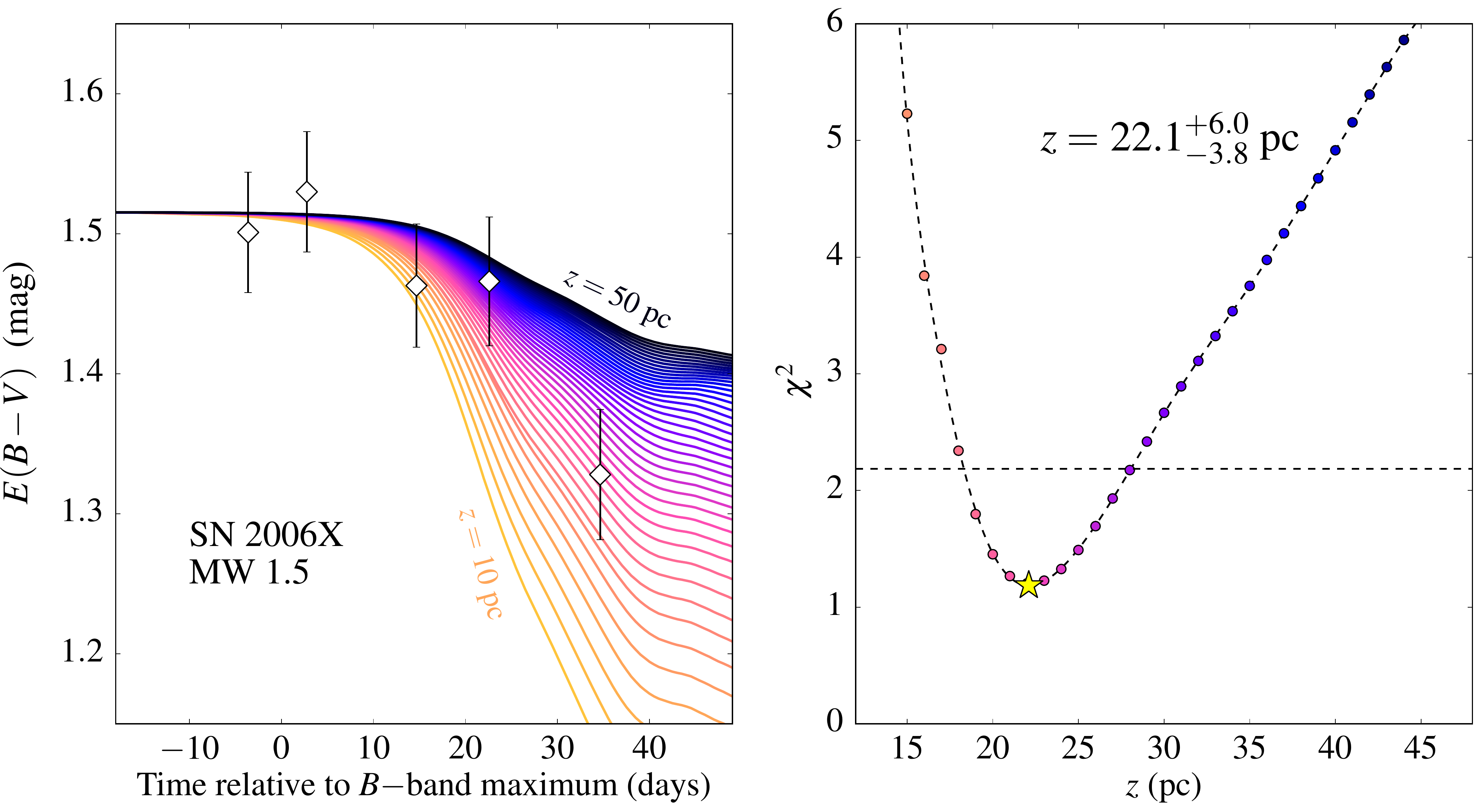}
\caption{Same as Fig.~\ref{fig:sn2006X_LMC} but for \mw type dust.}
\label{fig:sn2006X_MW}
\end{center}
\end{figure*}

In the left panel of Fig.~\ref{fig:sn2006X_MW}, we compare colour excess values for SN~2006X with models of \mw type dust located between 10 and 50~pc (step size of 1~pc). 
Similarly to what is found for \lmc type dust, \mw type dust models also predict a post-maximum drop in \ebv. In this case, however, the most-probable distance (see right panel of Fig.~\ref{fig:sn2006X_MW}) is lower and characterized by smaller uncertainties: \revised{\refereenew{$z=22.1^{+6.0}_{-3.8}$}~pc}. 

\begin{figure}
\begin{center}
\includegraphics[width=0.93\columnwidth]{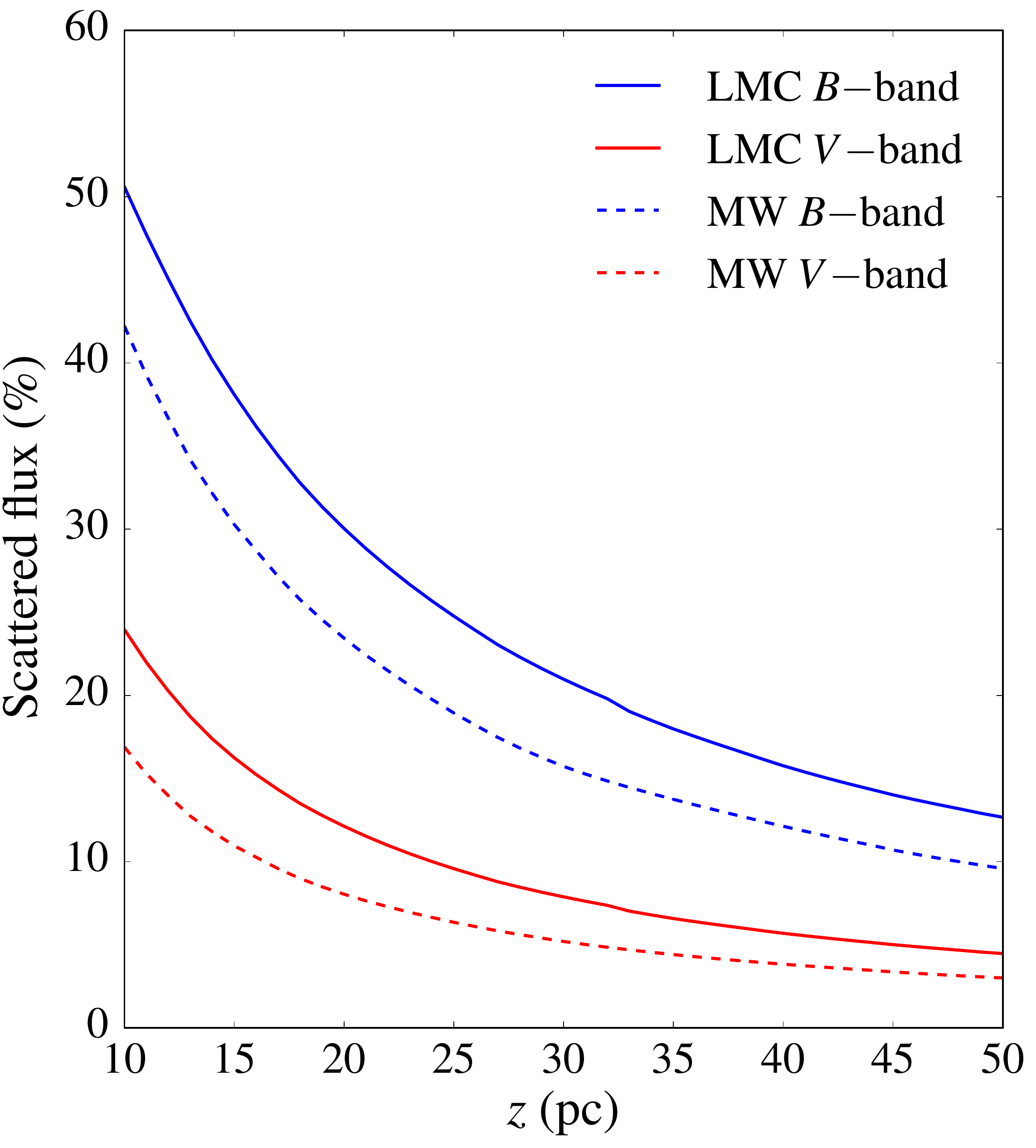}
\caption{Relative contribution of scattering \revised{photons} to the total flux in the $B-$ (blue lines) and $V-$band (red lines). Solid lines refer to \lmc type dust, while dashed lines to \mw type dust. Curves are calculated 35~d after $B-$band maximum, in correspondence to the last \ebv~value for SN~2006X (see Table~\ref{tab:sn2006x}).}
\label{fig:scattering}
\end{center}
\end{figure}

The reason why different dust compositions lead to different distance values and uncertainties is explained in Fig.~\ref{fig:scattering}, where we show the relative contribution of scattered \revised{photons} in the $B-$ and $V-$band as a function of dust distance $z$. The different changes in \ebv~between \lmc~and \mw type dust are directly related to the different number of scattered \revised{photons} contributing to the total flux. Due to the smaller albedo values of \mw~compared to \lmc type dust (see table~1 of \citealt{goobar2008}), the scattering contribution and thus the changes in \ebv~for a given $z$ are smaller in the former model. The best match to the observed \ebv~values is therefore found at lower $z$ for \mw~compared to \lmc type dust. In addition, Fig.~\ref{fig:scattering} shows that the scattering contribution decreases with dust distance and that this variation is steeper at lower $z$. Moving to increasingly smaller values of $z$, the change in \ebv~ is thus faster and the modelled curves of Figs.~\ref{fig:sn2006X_LMC} and \ref{fig:sn2006X_MW} less and less similar. As a consequence, the lower distance found in the \mw type dust model is also affected by smaller uncertainties.

Distance values derived above can be compared to estimates made for SN~2006X using different techniques. In particular, detections of light echoes in SN~2006X have been reported independently by \citet{wang2008b} and \citet{crotts2008} and used to constrain the dust distance under the single-scattering assumption. Based on Hubble Space Telescope (HST) data obtained 300~d post-maximum, \citet{wang2008b} concluded that a thick dust layer should be located somewhere between $\sim$~27 and 170~pc away from the SN site. Combining these same measurements with HST data obtained 680~d post-maximum, \citet{crotts2008} inferred a dust distance of $26.3\pm3.2$~pc. The latter is in excellent agreement with the values obtained from our modelling. Specifically, their estimate is consistent with predictions from our \lmc~and \mw type dust models within \revised{\refereenew{0.6} and \refereenew{0.2}}~$\sigma$, respectively.

\begin{table}
\centering
\caption{Average time and extracted \ebv~values for each phase interval of SN~2006X. Times and \ebv~values correspond to the white diamond points in the left panels of Figs.~\ref{fig:sn2006X_LMC} and \ref{fig:sn2006X_MW}.}
\label{tab:sn2006x}
\begin{normalsize}
\begin{tabular}{ccc}
\hline
Phase & $t_\mathrm{ave}$ & \ebv \\
(days from $M^\mathrm{obs}_{B,0}$) & (days from $M^\mathrm{obs}_{B,0}$) & (mag) \\
\hline
$[-10,0]$ & $-$3.6 & 1.501 $\pm$ 0.043 \\ 
$[0,+10]$ & $+$2.8 & 1.530 $\pm$ 0.043 \\ 
$[+10,+20]$ & $+$14.7 & 1.463 $\pm$ 0.044 \\ 
$[+20,+30]$ & $+$22.6 & 1.466 $\pm$ 0.046 \\ 
$[+30,+40]$ & $+$34.6 & 1.328 $\pm$ 0.046 \\ 
\hline
\end{tabular}
\end{normalsize}
\end{table}

As discussed in Section~\ref{sec:lcbroad}, the SN templates used to calculate modelled curves were stretched so that the resulting light curve after interaction with dust has $\Delta M^\mathrm{obs}_{B,15}$ equal to the one observed for SN~2006X, \mbox{$\Delta M^\mathrm{06X}_{B,15}=1.17\pm0.05$~mag} \citep{wang2008a}. As a by-product of our calculations, we can thus estimate the intrinsic decline rate of SN~2006X for the most-probable distance values derived above. An intrinsic decline rate of \mbox{$\Delta M^\mathrm{em}_{B,15}=1.20\pm0.01$~mag} is required for both \lmc type dust at \revised{\refereenew{$z=27.5^{+9.0}_{-4.9}$}~pc} and \mw type dust at \revised{\refereenew{$z=22.1^{+6.0}_{-3.8}$}~pc}. This decline rate is consistent with -- although more precise than -- the value derived by \citet{wang2008a} using the empirical relation proposed by \citet{phillips1999}, \mbox{$\Delta M^\mathrm{em}_{B,15}\sim\Delta M^\mathrm{obs}_{B,15}+0.1\times E(B-V)=1.31\pm0.05$}~mag.

\subsection{SN~2014J}
\label{sec:2014J}

SN~2014J was discovered in the nearby galaxy M82 and, owing to its proximity ($D\sim3.53$~Mpc), has been the subject of several studies in the past few years. SN~2014J was found to suffer significant extinction from the host galaxy, with an \textit{averaged} \ebv~value between $\sim$~1.2 and 1.4~mag \citep{amanullah2014,tsvetkov2014,foley2014,brown2015,amanullah2015,marion2015}. Here we use multi-band photometry from \citet{amanullah2015} to study the evolution of \ebv~with time. Six phase intervals are selected, and the \ebv~values for each interval are reported in Table~\ref{tab:sn2014j}. Unlike SN~2006X, SN~2014J does not show any clear evolution in the derived colour excess, with values extracted between $-$5 and +25~d compatible with a constant \ebv~$\sim$~1.3~mag. The lack of variability is consistent with the findings from \citet{brown2015} while in conflict with the time-evolution suggested by \citet{foley2014}.

\begin{table}
\centering
\caption{Same as Table~\ref{tab:sn2006x} but for SN~2014J. Times and \ebv~values correspond to the white diamond points in the left panels of Figs.~\ref{fig:sn2014J_LMC} and \ref{fig:sn2014J_MW}. }
\label{tab:sn2014j}
\begin{normalsize}
\begin{tabular}{ccc}
\hline
Phase & $t_\mathrm{ave}$ & \ebv \\
(days from $M^\mathrm{obs}_{B,0}$) & (days from $M^\mathrm{obs}_{B,0}$) & (mag) \\
\hline
$[-5,0]$ & $-$2.6 & 1.337 $\pm$ 0.027 \\ 
$[0,+5]$ & $+$1.1 & 1.313 $\pm$ 0.027 \\ 
$[+5,+10]$ & $+$8.1 & 1.315 $\pm$ 0.028 \\ 
$[+10,+15]$ & $+$17.6 & 1.314 $\pm$ 0.046 \\ 
$[+15,+20]$ & $+$21.2 & 1.337 $\pm$ 0.035 \\
$[+20,+25]$ & $+$29.0 & 1.290 $\pm$ 0.047 \\ 
\hline
\end{tabular}
\end{normalsize}
\end{table}

\begin{figure*}
\begin{center}
\includegraphics[width=1\textwidth]{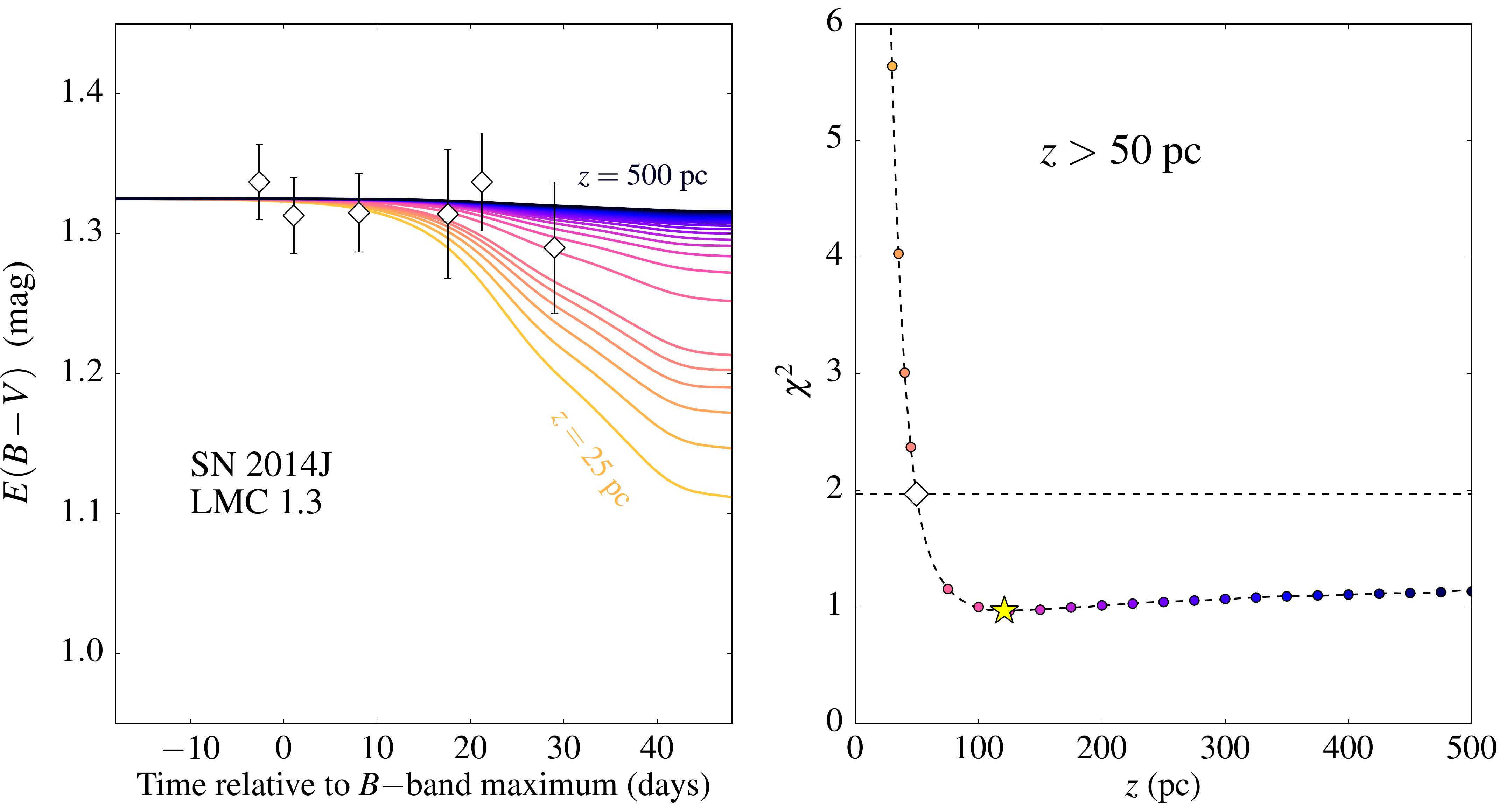}
\caption{\textit{Left panel.} \ebv~temporal evolution of SN~2014J (white diamonds) compared to model predictions for \lmc type dust with \ebvdlos~$=1.3$~mag. Models are calculated between 25 and \referee{75} pc (step size of 5~pc) and between 100 and 500~pc (step size of 25~pc). Model curves have been shifted by \referee{+$0.025$~mag to match the averaged value of the first two epochs of SN 2014J}. \textit{Right panel.} $\chi^2$ as a function of dust distances, where each point corresponds to a  curve in the left panel. A \refereenew{14}th-order polynomial fitting (dashed curve) is performed and used to constraint the most-probable value (yellow star) and a lower limit (white diamond) for the dust distance. The latter value is also reported in the panel.}
\label{fig:sn2014J_LMC}
\end{center}
\end{figure*}

The left panel of Fig.~\ref{fig:sn2014J_LMC} shows colour excess values estimated for SN~2014J, together with predicted curves from our \lmc type dust models. To encompass a larger distance range compared to that investigated for SN~2006X, here models are calculated between 25 and \referee{75} pc (step size of 5~pc) and between 100 and 500~pc (step size of 25~pc). \referee{We use models with \ebvdlos~$=1.3$~mag but apply a shift of +0.025 mag to the extracted curves to match the averaged value of the first two epochs of SN~2014J.}
As shown in the right panel of Fig.~\ref{fig:sn2014J_LMC}, the most-probable dust distance is constrained to be \revised{\refereenew{$z\sim121$}~pc}. Modelled curves at larger distances, however, tend to a constant \ebv~$=1.325$~mag and are all consistent with SN~2014J at the epochs investigated. Hence, we infer a 1$\sigma$ lower limit on the distance and conclude that dust towards SN~2014J is located at \revised{$z\gtrsim\referee{50}$~pc}.

\begin{figure*}
\begin{center}
\includegraphics[width=1\textwidth]{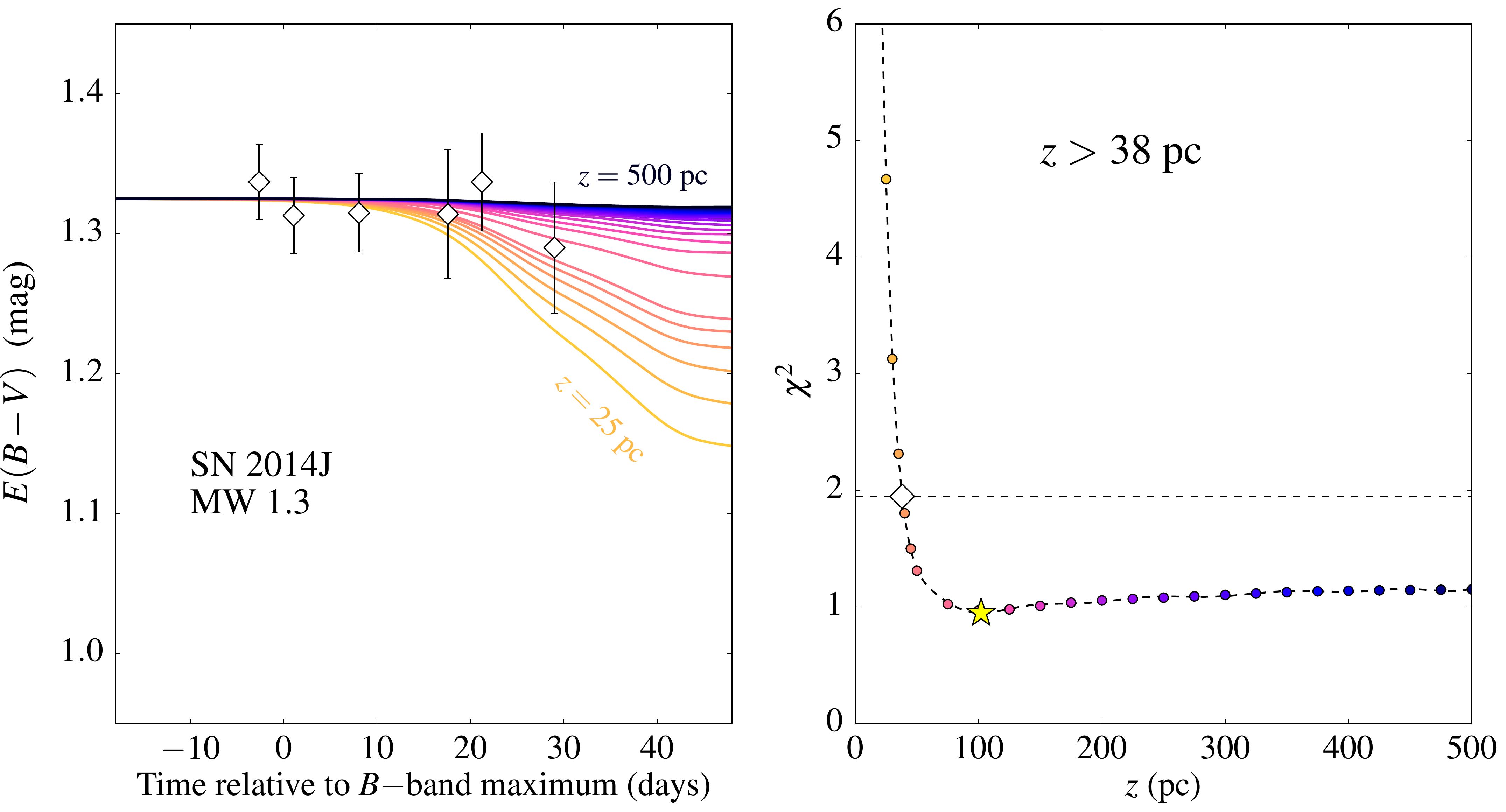}
\caption{Same as Fig.~\ref{fig:sn2014J_LMC} but for \mw type dust.}
\label{fig:sn2014J_MW}
\end{center}
\end{figure*}

Fig.~\ref{fig:sn2014J_MW} presents a comparison between \ebv~values estimated for SN~2014J and those predicted by \mw type dust models. Predicted curves are calculated for the same distance range of \lmc type models.
For the same reason discussed in Section~\ref{sec:2006X}, distance values estimated for \mw type dust are smaller than those for \lmc type dust. Specifically, we find that models with dust at \revised{$z\gtrsim\refereenew{38}$~pc} are consistent with data within 1$\sigma$ and infer a most-probable distance of \revised{\refereenew{$z\sim102$}~pc}.

Light echo detections towards SN~2014J were reported by \citet{crotts2015} and \citet{yang2017}. Using HST imaging data taken 216~d after $B-$band maximum, \citet{crotts2015} reported the detection of a light echo corresponding to dust at $\sim$~330~pc and of a possible inner one at $\sim$~80~pc. Combining these data with additional HST observations at +277, +365 and +416~d, \citet{yang2017} confirmed the presence of the outer echo for which, however, they inferred a smaller and more precise distance of $228\pm7$~pc. In addition, they detected a diffuse light echo indicative of dust located between $\sim$~100 and 500~pc away from the SN. Distance values inferred by both studies are compatible with our lower limits for \lmc~and \mw type dust models.

When calculating modelled \ebv~curves above, SN templates were stretched so that the observed decline rate after interaction with dust matches that inferred for SN~2014J. In particular, following \citet{galbany2016} we averaged four different estimates in the literature \citep{tsvetkov2014,ashall2014,kawabata2014,marion2015} and use \mbox{$\Delta M^\mathrm{14J}_{B,15}=1.06\pm0.06$}~mag. For both \lmc~and \mw type dust, lower limits derived above imply an 
intrinsic decline rate \referee{\mbox{$\Delta M^\mathrm{em}_{B,15}=1.07\pm0.01$}}~mag. In contrast, assuming \mbox{\ebv~$=1.3\pm0.1$}~mag, the empirical relation from \citet[][see Section~\ref{sec:2006X}]{phillips1999} would give an intrinsic decline rate of \mbox{$\Delta M^\mathrm{em}_{B,15}=1.19\pm0.06$}~mag. The two different values are consistent with each others, although our estimate is more precise and not relying on empirical observations.





\section{Discussion}
\label{sec:discussion}

In Section~\ref{sec:observations} we compared the \ebv~temporal evolution predicted by our \lmc~and \mw type dust models with that inferred for two highly-extinguished SNe~Ia, SN~2006X and SN~2014J. This comparison allowed us to place constraints on the dust location towards these SNe~Ia. For SN~2006X, we obtain distance values of \revised{\refereenew{$z=27.5^{+9.0}_{-4.9}$}~pc} and of \revised{\refereenew{$z=22.1^{+6.0}_{-3.8}$}~pc} for \lmc~and \mw type dust, respectively, in good agreement with previous estimates from light echo detections \citep[$26.3\pm3.2$~pc,][]{crotts2008}. For SN~2014J, instead, our models suggest that dust is located at distances larger than \revised{$\sim$~\referee{50} and \refereenew{38}~pc} for \lmc~and \mw type dust, respectively, again in good agreement with light echo measurements \citep{crotts2015,yang2017}.

As already discussed in Section~\ref{sec:lcbroad}, a constant \ebv~in the time window covered by the observations (from around maximum to about 30~d after) is not only consistent with fairly distant but also with very nearby dust. Therefore, the \ebv~time evolution derived for SN~2014J could also be consistent with dust with \ebvdlos~$\sim$~1.7~mag and located at radii smaller than $\sim$~10$^{16}$~cm ($\sim$~0.003~pc, see Fig.~\ref{fig:ebv}). Although we can not rule out this possibility, we find it rather unlikely since dust at these small distances should be either destroyed by the SN radiation \citep{amanullah2011} or produce  an infrared thermal emission that is not seen in SN~2014J and other normal SNe~Ia \citep{johansson2013,johansson2017}. We thus consider the very-distant dust case more plausible to explain the constant \ebv~behaviour observed in SN~2014J.

\referee{The dust region in our simulations is modelled as a single cloud with \ebvdlos~(see Section~\ref{sec:models}). In the case of multiple clouds with the same combined \ebvdlos, the results of this study would be unaffected. We note, however, that distance values estimated here would now be taken as representative of the outermost cloud (i.e. the more distant from the SN).}

Our findings suggest that, at least for the two SNe investigated, dust is likely to be located at IS-scale distances ($\sim$~10$-$100~pc). In the following, we briefly discuss how dust and gas at these scales might be compatible with other peculiar properties displayed by SNe~Ia suffering dust extinction.

\subsection{Peculiar extinction and polarization}
\label{sec:rvpol}

As discussed in Section~\ref{sec:introduction}, highly-reddened SNe~Ia are characterized by other peculiar properties. First, they are affected by unusually steep \textit{extinction} law, with inferred \mbox{$R_\mathrm{V}$} values significantly smaller than those typically observed for extinguished Galactic stars ($R_\mathrm{V}\sim3.1$). Secondly, they display a \textit{polarization} curve peaking at shorter wavelengths than those typically seen for extinguished stars in our Galaxy ($\sim$~0.55~$\mu$m). 

To account for \textit{both} these observational evidences, recent modelling efforts appear to favour an IS over a CS dust origin, provided that the dust grain sizes are significantly smaller than those found in the Galaxy \citep[see e.g.][]{gao2015,nozawa2016,hoang2017}.
The scenario recently proposed by \citet{hoang2017} holds promises for explaining why the environment around some SNe~Ia seems to prefer smaller dust grains. In this picture, a shift to smaller size distributions is believed to originate from cloud-cloud collisions induced by the SN radiation pressure. In particular, assuming that a number of different clouds are present along the DLOS, the more nearby clouds are first accelerated by the SN radiation pressure and later collides with more distant clouds. Cloud-cloud interactions are effective in destroying big grains, leading to the enhancement of grains smaller than $\sim$~0.1~$\mu$m and to the overall size distribution biased to smaller values.

The alignment of the polarization angle with the IS magnetic field of the host galaxy has been used as an additional strong argument favouring IS over CS dust scenarios \citep{patat2015,zelaya2017}. By modelling extinction and polarization data of the four SNe~Ia studied by \citet{patat2015}, \citet{hoang2017} finds that the observed alignment is compatible with that predicted by the radiative torque (RAT) mechanism (see \citealt{lazarian2015} for a recent review). In particular, the alignment observed in three SNe~Ia (SN~1986G, SN~2006X and SN~2014J) is consistent with that of our Galaxy -- where big grains are more efficiently aligned than small grains -- and ascribed to RAT induced by IS radiation fields. In contrast, modelling of SN~2008fp requires small grains to be aligned as efficiently as big grains. In the scenario described above, \citet{hoang2017} interprets this peculiar alignment as a consequence of RAT induced by the SN radiation itself, a picture that is found to work best for dust located at distances between $\sim$~1 and 20~pc.

Here we note that calculations presented in this work have been performed using \mw~and \lmc type dust. With these compositions, dust clouds located at distances inferred from Section~\ref{sec:observations} ($\gtrsim$~20~pc) are not able to reproduce the low $R_\mathrm{V}$ values observed in SNe~Ia. Using a more appropriate grain size distribution as the one predicted by \citet{hoang2017} would likely lead to more compatible $R_\mathrm{V}$ values at these distance scales. The different dust composition would also lead to different \ebv~evolution and it thus might affect the distance estimates here derived. From the comparison in Fig.~\ref{fig:ebv}, however, we expect a different dust composition to only marginally affect the \ebv~evolution and hence the estimated dust location.

To summarize, the grain size distribution and alignment function calculated by \citet{hoang2017} are indicative of dust located at IS-scales ($\gtrsim$~1~pc), in good agreement with
distances estimated in this work (Section~\ref{sec:observations}) and with those inferred from light echo detections towards some SNe~Ia.

\subsection{Variable blue-shifted absorptions}
\label{sec:linevar}

For the two SNe~Ia investigated in this study, variable, blue-shifted components have been detected in the Na\,{\sc i}~D and K\,{\sc i} absorption profiles. \revised{Variable sodium was observed in SN~2006X \citep{patat2007}, while SN~2014J showed variability in the potassium profile but not in the sodium one \citep{goobar2014,foley2014,ritchey2015,graham2015,jack2015,maeda2016}}. These variations were interpreted as arising from outflowing CS gas where sodium and potassium are subject to photoionization and later recombination effects \revised{\citep[][but see \citealt{maeda2016} for an alternative interpretation in the IS scenario]{patat2007,graham2015}}. To explain the observed variabilities, the gas was inferred to be at distances (0.1$-$1.0)$\times$10$^{17}$~cm ($\sim$~0.003$-$0.03~pc) in SN~2006X \citep{patat2007}, while (0.6$-$1.6)$\times$10$^{19}$~cm ($\sim$~2$-$5~pc) in SN~2014J \citep{graham2015}.

In order for these components to originate within IS clouds (i.e. at distances where dust in SN~2006X and SN~2014J appears to be located), one would need to invoke some geometric effect. For instance, the varying absorption features might stem from IS clouds whose covering factor changes with time due to the rapid expansion of the underlying SN photosphere. This possibility was in part ruled out by \citet{patat2010}, which found that the variations in the Na\,{\sc i}~D equivalent width ($\lesssim$~40~m\AA) induced by a fractal patchy IS medium are much smaller than those observed in some SNe~Ia (e.g. $\sim$~300~m\AA{} for SN~2006X). Large variations as those seen in SN~2006X, however, could still be consistent with geometric effects if the IS medium is highly non-homogeneous. For instance, \citet{chugai2008} suggested that the strongest variability observed in the Na\,{\sc i}~D profile of SN~2006X (``D'' component in \citealt{patat2007}) can be explained if the SN is \textit{first} seen through the hole of a cluster of clouds while \textit{later} partially occulted by the cluster. Using our Monte Carlo code, we verified that this particular distribution of the IS clouds can indeed reproduce the variability observed in SN~2006X.

A common argument against geometric effects in IS clouds relies on the fact that the observed blue-shifted components of Na\,{\sc i}~D, Ca\,{\sc ii}~H\&K and K\,{\sc i} do not vary simultaneously \citep{patat2007}. For instance, the blue-shifted components observed in the Ca\,{\sc ii}~H\&K profile of SN~2006X did not show any time evolution as was the case for the Na\,{\sc i}~D. In the CS scenario, instead, the lack of variability of calcium has been explained in terms of the higher ionization potential of calcium compared to sodium \citep{patat2007}. This argument has been challenged by \citet{chugai2008} who pointed out that this is also expected in the IS scenario provided that the variable sodium components arise from clouds where calcium is depleted on dust grains.

Another issue for the IS scenario is to explain why the variable components are preferentially seen blue-shifted. A possible answer is found in the scenario proposed by \citet{hoang2017}, where small grain sizes inferred from extinction and polarization analysis are due to cloud-cloud collisions induce by SN radiation (see Section~\ref{sec:rvpol}). In this picture, the SN radiation pressure accelerates the more nearby clouds to move \textit{towards} the observer and to later collide with the more distant clouds. Absorbing gas in these moving clouds would thus be detected via blue-shifted components. In addition, we note that the scenario of \citet{hoang2017} can also account for the observational preponderance of variable Na\,{\sc i}~D and K\,{\sc i} features in SNe~Ia with large dust extinction \citep{patat2007,blondin2009,phillips2013,graham2015}. In fact, cloud-cloud collisions would not only destroy big grains to smaller sizes (see Section~\ref{sec:rvpol}) but also lead to an enhancement in the gaseous phase of atoms such as sodium, calcium and potassium. Finally, this scenario might also explain why the SNe~Ia with unusually high sodium absorption in the sample studied by \citet{phillips2013} display blue-shifted Na\,{\sc i}~D absorptions.


In summary, although variable blue-shifted components have been historically explained by CS gas around some SNe~Ia, growing evidences suggest that these can also be consistent with geometric effects arising at IS-scale distances.

\section{Conclusions}
\label{sec:conclusions}

We have presented a new approach to calculate the location of dust in reddened SNe~Ia. This technique exploits the fact that dust regions at different locations produce distinct behaviours on the observed SN light curve \citep{amanullah2011}. Using Monte Carlo calculations, we have shown that the detailed evolution of the colour excess \ebv~provides strong constraints on the dust distance from the SN. Although this approach is complementary to distance estimates from light echo measurements, unlike the latter it does not require to obtain late-time HST imaging but only photometric data at relatively earlier epochs. Moreover, the use of Monte Carlo methods naturally incorporates the effect of multiple scattering on dust particles which is not taken into account when estimating distances from light echo measurements.

We applied our approach to two well-studied SNe~Ia suffering large extinction: SN~2006X and SN~2014J. In the case of SN~2006X, our models predict SN-dust distances of \revised{\mbox{\refereenew{$z=27.5^{+9.0}_{-4.9}$}}} and \revised{\mbox{\refereenew{$z=22.1^{+6.0}_{-3.8}$}~pc}} depending on whether \lmc~or \mw type dust is used. These values are in excellent agreement with predictions from light echo detections \citep[$26.3\pm3.2$~pc,][]{crotts2008}. In the case of SN~2014J, \ebv~is found to be constant in the time window investigated, allowing us to place only lower limits on the dust distance. Specifically, dust is constrained to be located at distances larger than \revised{$\sim$~\referee{50} and \refereenew{38}~pc} for \lmc~and \mw type dust, respectively, in good agreement with light echo suggestions \citep{crotts2015,yang2017}.

Our study indicates that for the two SNe~Ia investigated dust is likely to be located at IS-scale distances. \referee{This is in conflict with findings from \citet{cikota2017}, which support multiple scattering in a local environment for these SNe, but compatible with the non-detection of infrared thermal emission in SN~2014J \citep{johansson2017}}. We have discussed how our results could be consistent with other peculiar properties displayed by some highly-reddened SNe~Ia. First, their peculiar extinction and polarization properties could be ascribed to small IS grains produced by cloud-cloud collisions induced by the SN radiation pressure, as proposed by \citet{hoang2017}. Secondly, the variable, blue-shifted components observed in Na\,{\sc i}~D and K\,{\sc i} features of some highly-reddened SNe~Ia (including SN~2006X and SN~2014J) could be a consequence of the SN photosphere expanding in a non-homogeneous IS medium \citep{chugai2008}. In particular, we confirmed this possibility with our Monte Carlo code by exploring a similar geometry to the one proposed by \citet{chugai2008}. Finally, the cloud-cloud collision scenario proposed by \citet{hoang2017} might explain why SNe~Ia with unusually strong Na\,{\sc i}~D absorption are characterized by blueshifted Na\,{\sc i}~D components \citep{phillips2013} and why variabilities in the latter are preferentially seen in SNe~Ia suffering large extinction.

While in this study we focused on the evolution of \ebv, in the future we will investigate the effect of dust on the extinction in different bands $A_X$ and thus on the colour excess $E(X-V)$. Combining information from different wavelengths will help provide even more stringent constraints on the dust location. Also, since scattering and absorption properties have not been computed for the grain distribution predicted by \citet{hoang2017}, in this study we have restricted to using \lmc~and \mw type dust composition. As a result, our models predict $R_\mathrm{V}$ values consistent with those observed in the Galaxy (and thus inconsistent with SNe~Ia) at the pc-scale distances inferred from the \ebv~evolution. In the future, we plan to perform Monte Carlo calculations using dust properties inferred from \citet{hoang2017}. When placing dust at IS distances, we expect these calculations to give $R_\mathrm{V}$ values consistent with those seen in SNe~Ia, while only marginally affecting the \ebv~time evolution and thus distance estimates derived in this study. 

Although in this work we focused on presenting and testing our technique on two-well known SNe~Ia, in the future we plan to apply the same approach to a larger sample of highly-reddened SNe~Ia. In particular, the first year of the upcoming wide-field ground-based Zwicky Transient Facility \citep[ZTF,][]{kulkarni2016} survey will provide us with roughly 100 (40) SNe~Ia with host \ebv~$\geq$~0.4 (0.5)~mag, 70 (25) of which will be discovered before maximum light. This will allow us to estimate colour excess evolutions and thus dust distance locations for a statistically-significant sample, and possibly help solve the CS vs IS long-standing debate.

\section*{Acknowledgements}

\revised{\referee{The authors are thankful to the anonymous reviewer for his/her valuable comments.} The authors acknowledge support from the Swedish Research Council (Vetenskapsr\aa det) and the Swedish National Space Board.}

\bibliographystyle{mn2e}
\bibliography{bulla2017a}

\end{document}